# Generation of antitropic electron beams by self-generated electric field. Kinetic description

N. S. Stepanov[1], L. A. Zelekson[2]

[1] Nizhny Novgorod State University, Russia. E-mail: Stepanov@rf.unn.ru

[2] Nizhny Novgorod State Conservatory, Russia. E-mail: lionhope@gmail.com

Abstract

This paper makes use of a one-dimensional kinetic model to investigate the nonlinear longitudinal dynamics of electron beams generated in the plasma under the influence of a self-generated electric field. It is proved that the correct solution of the Vlasov equation is characterized by sudden change in the resonant part of the distribution function. Hence, in particular, the incorrectness of the established in the literature point of view follows that the fast, with velocities over the wave velocity $V$, and slow, with velocities under $V$, trapped particles are described by the same distribution function. Also for the first time it is shown that the self-generated strong electric field always produces antitropic electron beams with the velocities much larger than the value $V$, including the cold plasma limit. The possibility of implementing a new class of self-consistent wave structures with a nonzero average potential is shown. Maximum possible amplitude of a self-generated electric field in a stationary collisionless plasma is determined.

## I. INTRODUCTION

High-energy accelerators and transport systems have a wide variety of applications ranging from basic research in high energy and nuclear physics, to applications such as medical physics and heavy ion fusion. Experiments on plasma acceleration by laser and particle beam have shown a major role of nonlinear dynamics of intense charged particle acceleration process. For the first time, such processes are investigated more fully by Bernstein, Greene, and Kruskal (BGK, Ref. [1]). BGK solutions, however, have a fundamental flaw — they do not satisfy the uniqueness condition. Subsequent authors, applying the BGK technique to a number of specific tasks, consider that the polysemy problem is removed by a priori completions of the distribution function (see, e.g., Refs. [2] and [3]) and references therein). It is assumed that the fast and slow trapped particles are described by the same function. However, as shown below, this assumption contradicts the very principle of self-consistency of the original equations. In particular, solutions of other authors are not satisfied the boundary conditions of the Vlasov equation in its differential form.

In this paper, we found the unambiguous stationary solution of the one-dimensional (1D) Vlasov equation for collisionless plasma without any priori completions of the distribution function. It is proved that it is characterized by sudden change in its resonant part. Besides, self-generated strong electric field always produces antitropic electron beams with the velocities much larger than the value $V$, including the cold plasma limit. The possibility of implementing a new class of self-consistent wave structures is shown.

## II. KINETIC EQUATION AND ITS SOLUTION

As in Ref. [1], we consider the class of solutions, when the electric field of the wave $E(z,t)$, and the distribution function of electrons and ions are functions of the variable $\xi = z - Vt$, where $z$ is the Cartesian coordinate, $t$ is the time; wave velocity $V$ ($0 < V < \infty$) is assumed to be constant. Introducing wave potential $\varphi(\xi)$, so that $E(\xi) = -\partial \varphi / \partial z = -d\varphi/d\xi$, the original equations, i.e., the 1D non-relativistic kinetic equations for electrons and ions, can be written as

$$(v-V)\frac{\partial f_e}{\partial \xi} + \frac{e}{m_e}\frac{d\varphi}{d\xi}\frac{\partial f_e}{\partial v} = 0, \qquad (1)$$

$$(v-V)\frac{\partial f_i}{\partial \xi} - \frac{e}{m_i}\frac{d\varphi}{d\xi}\frac{\partial f_i}{\partial v} = 0. \qquad (2)$$

Here $f_{e,i}(v,x)$ is the 1D (i.e., integrated over the transverse velocity components) distribution function, $m_e$ and $m_i$ are electron and ion mass with charges $-e$ and $+e$, respectively, $e > 0$. Since Eqs. (1) and (2) are similar, further in the expressions relating only to electrons, we will omit the symbol $e$. In the expressions relating to ions, it is sufficient to replace $e/m_e$ with $-e/m_i$, and so on.

As is known, the general solution of Eq. (1) is an arbitrary function of its characteristic variable, which we define as follows:

$$u(v,\psi) = \sqrt{(v-V)^2 - \psi V^2}, \qquad (3)$$

where $\psi = 2e\varphi/(mV^2)$ is the dimensionless wave potential. In the velocity interval

$$v_1 \leq v \leq v_2, \ v_{1,2}(\psi) = V(1 \mp \sqrt{\psi}), \ \psi \geq 0, \qquad (4)$$

when $u^2(v,\psi) < 0$ and the values $u(v,\psi)$ are imaginary [elliptic type of Eq. (1)], the electrons perform a finite motion in a wave potential well (further symbol $tr$ corresponds to trapped electrons). When $u^2(v,\psi) \geq 0$ (hyperbolic type Eq. (1) electrons are not trapped by the wave, i.e., passing electrons — symbol $p$.

With up to notation Eqs. (1)—(4) are naturally present in the papers of the authors mentioned above. However, the main attention was paid in them not to the kinetic equations, but to the study of the Poisson equation and the solving of arising at the same time, as noted above, the problem of the polysemy of its solutions. Now we will show that this problem is completely eliminated at the stage of analysis of kinetic Eqs. (1) and (2), without any a priori assumptions about the nature of the trapped particle distribution function and without selection of their parameters.

## A. Boundary conditions for function $f(v, \xi)$

We emphasize that the only singular point of Eq. (1) is the speed $v = V$, when the first term, and hence the second vanishes. This means the first velocity derivative of function $f(v, \psi)$ is zero when the synchronism $v = V$ occurs:

$$\left.\frac{\partial f[v, \psi(\xi)]}{\partial v}\right|_{v=V} = 0. \qquad (5)$$

As for the second factor in the first term, namely $\partial f[v, \psi(\xi)]/\partial \xi$, when $v = V$, its value and, hence, $f(v, \psi)$, can in principle be arbitrary (except, of course, the trivial case of a constant potential for any $\xi$). Thus, in stationary collisionless plasma during the transition through the synchronism function $f(v, \psi)$ discontinuity is allowed:

$$f(V+0, \psi) \neq (V-0, \psi). \qquad (6)$$

In contrast to the synchronism, the points of the surfaces $v = v_{1,2}(\psi)$ are not specific to Eq. (1). Therefore, when passing through them, in spite of the qualitative change in the nature of the motion of the electrons, the function $f(v, \psi)$ and its first velocity derivative are continuous for all wave potential values out of the plane $v = V$, including on the boundary (4):

$$f(v_{1,2}+0, \psi) = f(v_{1,2}-0, \psi), \quad \left.\frac{\partial f(v, \psi)}{\partial v}\right|_{v=v_{1,2}+0} = \left.\frac{\partial f(v, \psi)}{\partial v}\right|_{v=v_{1,2}-0}. \qquad (7)$$

Equations (7) can be considered as the boundary conditions for the function $f(v, \psi)$ when $v = v_{1,2}(\psi)$. And if the first expression in Eq. (7) one way or the other (mostly in integrated form[1]) appeared in the papers cited above, the second had never been mentioned.

## B. Algorithm for obtaining Vlasov equation solution

It is worth mentioning here the obvious requirement for the Vlasov equation solution from the physical point of view: it has to be real and non-negative for arbitrary values of $v$ and $\psi$. If we directly use the characteristic variable (3) as the function $f(v, \psi)$ argument, then this requirement is not feasible.

At the same time, the quadratic form of the variable $u(v, \psi)$

$$w(v, \psi) = [V + au(v, \psi)][V + a^* u^*(v, \psi)] \qquad (8)$$

---

[1] Note, incidentally, that using in these papers the conservation laws in integral form, as noted in Ref. [4], is a necessary condition but not sufficient, in contrast to their differential form.

[where *a* is an arbitrary complex constant and the symbol * denotes the complex conjugate, which is also a general integral of Eq. (1)], always is the real and positive definite, including the velocity interval (4). Therefore, unlike $u(v,\psi)$ we will use Eq. (8) as an argument of the general solution of Eq. (1):

$$f(v,\psi) = f[w(v,\psi)]. \tag{9}$$

In view of Eqs. (7) and (9) for the distribution function $f[w(v,\psi)]$ we obtain the relations $w(v,\psi)$:

$$w(v_1,\psi) = w(v_2,\psi), \quad \left.\frac{\partial w(v,\psi)}{\partial v}\right|_{v=v_1} = \left.\frac{\partial w(v,\psi)}{\partial v}\right|_{v=v_2}. \tag{10}$$

Substituting Eq. (8) into Eq. (10), it is easy to determine the constant *a*:

$$a_p^{(\pm)} = \pm 1, \quad a_{tr}^{(\pm)} = \pm i, \tag{11}$$

where the minus sign corresponds to slow ($v < V$) electrons, and the plus sign to fast ones ($v \geq V$). The discontinuity of parameter *a* when passing through the synchronism $v = V$ confirms the validity of Eq. (6).

With the equation (11) we obtain that, depending on the type of electrons, and therefore in different velocity *v* intervals, expressions for the argument $w(v,\psi)$ are different:

$$w(v,\psi) = w_p^{(+)}(v,\psi) \, [U(\psi)U(v-v_2)+U(-\psi)U(v-V)] +$$
$$+ w_p^{(-)}(v,\psi) \, [U(\psi)U(v_1-v)+U(-\psi)U(V-v)] + \tag{12}$$
$$+ [w_{tr}^{(+)}(v,\psi) \, U(v-V)U(v_2-v) + w_{tr}^{(-)}(v,\psi) \, U(V-v)U(v-v_1)] \, U(\psi),$$

where

$$w_p^{(\pm)}(v,\psi) = [V \pm u(v,\psi)]^2, \tag{13}$$

$$w_{tr}^{(\pm)}(v,\psi) = [V \mp \sqrt{-u^2(v,\psi)}]^2. \tag{14}$$

Hereinafter, $U(x)$ is the Heaviside step function: $U(x) = 1$ when $x \geq 0$ and $U(x) = 0$ when $x < 0$. It is easy to show that, Eq. (12) multiplied by $m_e/2$ coincides with the electron energy density in the wave field.

Firstly, it is important to note that, as follows from Eq. (14), this value for trapped fast electrons at any wave amplitude as opposed to passing ones is always less than for the trapped slow particles[1]: $w_{tr}^{(+)}(v,\psi) < w_{tr}^{(-)}(v,\psi)$. In particular, it holds for the synchronism:

$$w_{tr}^{(+)}(V,\psi) - w_{tr}^{(-)}(V,\psi) = -4V\sqrt{\psi}. \tag{15}$$

Secondly, since the Eqs. (8)—(14) have not the electron temperature, they are valid, including in the case of the cold plasma.

---

[1] In this connection, we note similar results obtained in Refs. [5] and [6], which relate to the energy density of the fast and slow waves in moving media.

Specifically dependence $f[w(v,\psi)]$, in turn, is determined by its structure at an arbitrary point $\xi$. So, with $\xi \to \infty$ the potential $\varphi(\xi)$ tends to zero and therefore the plasma becomes equilibrium, then the function $f[w(v, 0)]$ must describe the unperturbed electron distribution $F(v^2)$:

$$f[w(v, 0)] = F(v^2). \tag{16}$$

Thus, the final expression for $f(v,\varphi)$ can be obtained by substituting the right side of Eq. (16) instead of the velocity $v$ argument $\sqrt{w(v,\psi)}$.

As the equilibrium, we use below the Maxwellian distribution:

$$F(v^2) = \frac{n_0}{\sqrt{2\pi}v_e}\exp(-\frac{v^2}{2v_e^2}), \tag{17}$$

where $v_e = \sqrt{\chi T_e/m_e}$ is the average electron thermal velocity, $T_e$ is their temperature, $\chi$ is the Boltzmann constant, $n_0$ is unperturbed electron density.

### C. Examination of distribution function $f(v, \xi)$

So, we get a self-consistent distribution function that takes into account both passing and trapped particles. We can say that its component describing the trapped particles is an analytic continuation in the case of a complex characteristic variable.

First, we note that, in accordance with the Eqs. (6), (14), and (15), at an arbitrarily small wave potential $\psi$ when passing through the synchronism plane, the function $f(v,\psi)$ always has a discontinuity. This fact makes the assumption of the constancy of the trapped electron distribution, which is the basis of many works, insolvent.

It follows from Eq. (15), as compared with the value at the boundaries of the potential well of the distribution function $v = v_{1,2}$ discontinuity is small only in two limiting cases: (a) $|\psi|<<(v_e/V)^4$; and (b) $|\psi|>>1$. The first of these corresponds to a fast weakly nonlinear wave, and the second one corresponds to a slow wave of high amplitude. In both cases, the number of electrons trapped near the synchronism plane is very small.

Qualitative distribution (17) is illustrated in Figs. 1 and 2 for several values of velocity and wave potential. Thus, with increasing wave potential, the domain of existence of both passing slow $(v<v_1)$, and fast $(v \geq v_2)$ electrons decreases, and the recession of decreasing branch of distribution function, which corresponds to the trapped fast electrons, becomes more significant.

In the case of a moderately strong waves $(|\psi|<1)$ with increasing $v$, the function $f^{(+)}(v,\psi)$ decreases monotonically, with its magnitude at the synchronism exceeds its value at the edge of the potential well: $f^{(+)}(V,y) > f^{(+)}(v_2,\psi)$. With increasing wave potential $(|\psi|\geq 1)$, the value $f^{(+)}(v,\psi)$, as well as $f^{(-)}(V,\psi)$, is reduced and the function $f_{tr}^{(+)}(v,\psi)$ has a maximum. The more $|\psi|$, the closer the extremum to the edge of potential well $v = v_2$. If $|\psi| \geq 1$, we have the inverse inequality $f^{(+)}(V,\psi) \leq f^{(+)}(v_2,\psi)$. As for the "passing" branch of distribution $f^{(+)}(v,\psi)$, its behavior is similar to this function $f_p^{(-)}(v,\psi)$: with increasing $v$, the larger $\psi$ the faster it decreases exponentially.

Thus, with increasing $\psi$ the number of both slow and fast passing electrons reduces and the number of trapped ones increases. Importantly, as the calculations show and can be seen clearly in Figs. 1 and 2, at any value $\psi$ fast trapped particles dominate unlike the slow ones, as the area limited by the curve $f_{tr}^{(+)}(v,\psi)$, by definition, equal to their number, is always larger than the area limited by the curve $f_{tr}^{(-)}(v,\psi)$. This prevalence becomes more prominent with increasing wave potential. In case of strong ($|\psi|>1$) nonlinearity, as seen in Figs. 1 and 2, the plasma is divided into two oncoming beams of fast trapped and slow passing electrons. If $\psi >> 1$, their velocities are equal respectively $v_2 = -v_1 = |\psi|^{1/2} >> V$.

Let us note another characteristic feature of the solution. Due to the exponential dependence (17), the velocity space of the prevalent electrons is determined by the inequality $w(v,\psi)/v_e^2 \leq 1$. In accordance with it, when $(V/v_e)^2 \leq 1$, the value of $w(v,\psi)$ can vary widely and accordingly, velocity space, that is, the distribution function is an area with a smooth surface (see Fig. 1). With increasing the parameter $V/v_e$, the possible size of values $w(v,\psi)$ is narrowed, resulting in irregularity function $f(v,\psi)$ (Fig. 2).

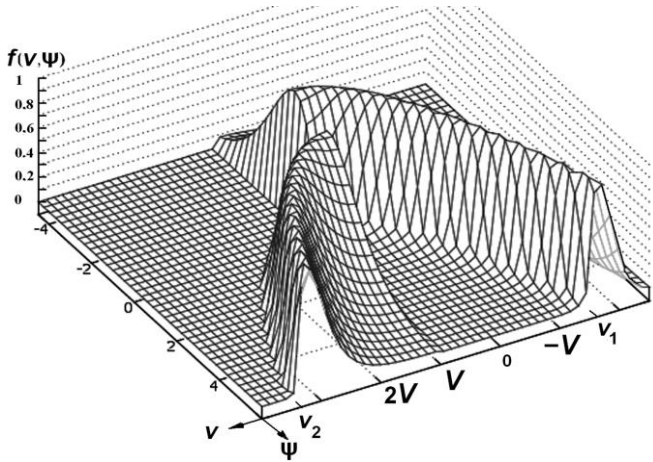 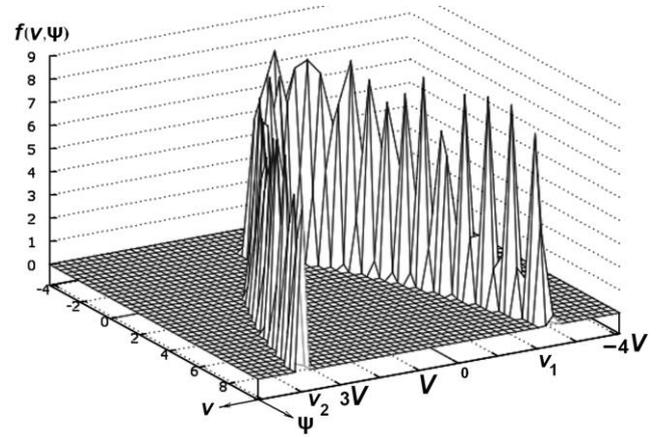

FIG. 1. The distribution $f(v,\psi)$: $v_e/V = 2$.   FIG. 2. The distribution $f(v,\psi)$: $v_e/V = 9$.

Characteristically, the situation does not change in principle, and the same irregularity arises if instead of Maxwellian distribution as an equilibrium one use the Heaviside step function

$$F(v^2) = \frac{n_0}{2v_e} U(v_e^2 - v^2) \qquad (18)$$

with the argument $\sqrt{w(v,\psi)}$ instead of $v$. Calculations show that for both functions at the same parameters graphs of $f(v,\psi)$ are virtually identical except at its maximum smoothness $(V/v_e)^2 \leq 1$. This fact we use further.

### III. SOLUTION OF POISSON EQUATION

Now, to find unambiguous electron and ion distributions, we consider the Poisson equation. In our notation, it takes the following form:

$$\frac{d^2\psi}{d\xi^2} + \frac{2\omega_e^2}{n_0 V^2}(n_i - n_e) = 0. \qquad (19)$$

Here $n_e$ and $n_i$ are the concentration of electrons and ions respectively:

$$n_{e,i}(\xi) = \int_{-\infty}^{+\infty} f_{e,i}(v,\psi)dv \qquad (20)$$

and $\omega_e = \sqrt{4\pi e^2 n_0/m_e}$ is the Langmuir frequency. Multiplying Eq. (19) by $d\psi/d\xi$ and integrating, we convert it to an equation describing the motion of a material particle in a potential field $\Phi(v,\psi)$:

$$\left(\frac{d\psi}{d\xi}\right)^2 = \left(\frac{2\omega_e}{V}\right)^2 [C - \Phi(\psi)], \qquad (21)$$

where

$$\Phi(v,\psi) = \Phi_i(V,\psi) - \Phi_e(V,\psi) = \int [n_i(V,\psi) - n_e(V,\psi)]d\psi,$$

$C = (E_0 V)^2/(4\pi n_0)$ is the constant of integration: $\Phi_{min} \leq C \leq \Phi_{min}$, $\Phi_{max}$ and $\Phi_{min}$ are the extremes of the potential function; $E_0$ is the maximum intensity of wave electric field.

As the integrand in Eq. (20) we use Eq. (18). This approach is justified by the fact that the characteristics of the integrand in Eq. (20) do not in principle affect the results, and most importantly, in terms of getting the final expression in the analytical form, without which it is difficult to identify possible new effects. Naturally, the numerical data are not absolute and may vary for other kind of equilibrium distribution. Nevertheless, they give an idea of the order of magnitude.

Substituting function (18) into Eq. (20), by integrating with Eqs. (12)—(14) we obtain:

$$\Phi_e(V,\psi) = \frac{1}{3}[B_+^{3/2} \mp B_-^{3/2} \pm A_-^{3/2} U(A_-) - A_+^{3/2} U(A_+)]. \qquad (22)$$

where

$$A_\pm(V,\psi) = \psi - (1 \pm v_e/V)^2, \quad B_\pm(V,\psi) = \psi + (1 \pm v_e/V)^2;$$

the above sign is used when $V \geq v_e$, the below one corresponds to $V < v_e$. Note that the last two terms in Eq. (22) owe their existence to trapped electrons. To be specific, we assume $m_e/m_i \approx 1/1800 \ll 1$, that allows to neglect the influence of the trapped ions. Then the function $\Phi_i(V,\psi)$ up to sign get the expression, whose structure is similar to the first two terms in Eq. (22).

As is known, the determining factor in the dynamics of solutions of Eq. (21) is the presence of stable [we denote them $\psi_{min}(V)$] and unstable [$\psi_{max}(V)$] equilibrium points that are defined by the extremes of the potential function: $\Phi_{max}=\Phi(\psi_{max})$ and $\Phi_{min}=\Phi(\psi_{min})$. At these points, the concentrations of electrons and ions are equal. This situation can be qualitatively different, depending on the value of $C$, which characterizes the degree of the wave nonlinearity.

Thus, in case of small wave amplitude ($|\psi| \ll 1$, that is, when $C - \Phi_{min} \ll \Phi_{max}$) taking the potential function $\Phi(v,\psi)$ of the Taylor series near the stable equilibrium point, we obtain $\Phi(\psi) = k^2(V)(\psi - \psi_{min})^2$ up to quadratic terms. In this expression,

$$k^2(V) = 2\left(\frac{\omega_e}{V}\right)^2 \frac{d^2\Phi(\psi_{min})}{d\psi^2} > 0 \qquad (23)$$

is the dispersion relation of weakly nonlinear stationary waves. What is important is the fact that if $\psi_{min} \neq 0$, their period-averaged potential $\varphi_{av} = m_e \psi_{max} V^2/(2e)$ is also not zero, in contrast, for example, to conventional Langmuir waves.

With a strong nonlinear ($|\psi| \geq 1$, i.e., when $C \leq \Phi_{max}$), finite solutions are known to be solitary waves (solitons). These parameters can be assessed with the following qualitative considerations. In the first place, their potential can vary from $\psi_{max}$ to $\psi_0$, where $\psi_c$ is the root of the equation $\Phi_{max} = \Phi(\psi_0)$. Secondly, according to Eq. (21) the amplitude $E_0$ has the order of value

$$E_0 = m_e \omega_e V (\Phi_{max} - \Phi_{min})^{1/2} / e. \qquad (24)$$

In turn, the values $\psi_{max}$, $\psi_0$, and $E_0$ allow us to estimate a characteristic width $\Delta$ of the soliton: $\Delta \sim (\psi_0 - \psi_{max})/E_0$. As a result we get for the $\Delta$:

$$\Delta = e(\psi_0 - \psi_{max}) / [m_e \omega_e V (\Phi_{max} - \Phi_{min})^{1/2}]. \qquad (25)$$

As the analysis of Eq. (21), it is necessary to distinguish between two cases: (a) when the wave velocity is much larger than the thermal electron velocity; or (b) comparable to it.

Thus, when $(V/v_e)^2 \gg 1$ (more specifically even when $V > 5 v_e$) there are two stable points, $\psi = 0$ and $\psi_{min} = 4.18$, and unstable one $\psi_{max} = 1$ (see Fig. 3). It is essential that their values are constant for velocities $V/v_e > 5$. Usually, excluding the trapped electrons, there is only one equilibrium state $\psi = 0$, corresponding to the usual Langmuir oscillations with a frequency $\omega_e$. With the rise of the second stable state the wave pattern changes significantly: oscillations with $\varphi_{av} = m_e \psi_{min} V^2/(2e) \neq 0$ can also exist in plasma.

For low amplitudes of these waves (i.e., $|\psi - \psi_{min}| \ll \psi_{min}$) using Eqs. (22) and (23) we obtain the following dispersion relation:

$$\omega(V) = \omega_e[0.5 + 0/2(v_e/V)^2]. \quad (26)$$

Thus, this wave frequency is comparable to $\omega_e$. As for their group velocity $v_{gr} = d\omega/dk$, then it follows from Eq. (26), it is much smaller than the value of $v_e$: $v_{gr} = 0.8 v_e^2/V$, as for the Langmuir waves.

Because of the trapped electrons, that is, the emergence of point $\psi_{min}$ in plasma with conventional Langmuir soliton is possible one with the potential changing from the value $\psi_{max} = 1$ at the edges (when $\xi \to \pm\infty$) up to a maximum $\psi_0 = 12$. Its amplitude and width can be calculated by substituting into Eqs. (24) and (25) the values $\psi_0 = 12$, $\Phi_{max} = -3606.2$ and $\Phi_{min} = -3604$ taken from Fig. 3.

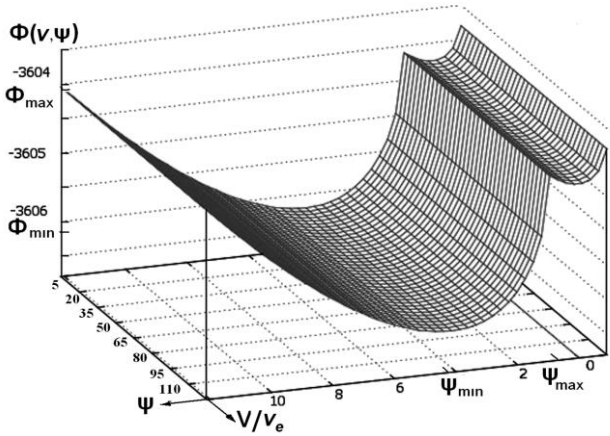 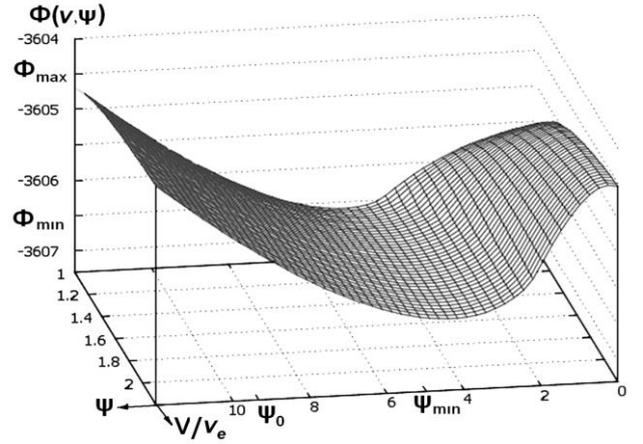

FIG. 3. The potential function $\Phi(V, \psi)$: $v_e/V > 5.2$.      FIG. 4. The potential function $\Phi(V, \psi)$: $0.9 < v_e/V < 2.2$.

Taking into consideration the trapped electrons leads to another effect: if from the potential (22) we remove corresponding to them terms, then, according to calculations, the maximum amplitude of the Langmuir soliton is of much greater size than it follows from Eq. (24).

Finally, we note that the results obtained here for the large parameter $(v_e/V)^2 \gg 1$, may also be obtained by applying the stationary phase method to Eq. (20), if we use the Maxwellian distribution (17) as the integrand.

With the wave velocity decrease, when it is in the range of $0.9 < v_e/V < 2.2$, the point $\psi = 0$ becomes unstable (i.e., $\psi_{max} = 0$) and there is only one stable equilibrium state $\psi_{min} = 5.0$ (see Fig. 4). In other words, in this case the plasma waves with zero average potential are absent and there are waves with $\varphi_{av} = 5m_e V^2/(2e)$. In the weakly nonlinear approximation of Eqs. (22) and (23) we find an expression for the frequencies:

$$\omega(k) = 0.4\omega_e[1 + \omega_e/(kv_e)], \quad kv_e \gg \omega_e.$$

As for the high-velocity wave, its frequency is comparable with Langmuir one and the group velocity $v_{gr} = -0.4\omega_e^2/k^2 v_e$ is negative and much less than the average thermal electron velocity.

In this velocity interval $0.9 < v_e / V < 2.2$ soliton parameters are as follows: its potential changes from the value of $\psi = 0$ at the edges (when $\xi \to \pm\infty$) up to the maximum $\psi_c = 9.6$. Its amplitude and width can be calculated by substituting into Eqs. (24) and (25) the values $\psi_{max} = 0$, $\psi_0 = 9.6$, $\Phi_{max} = -3605.6$, and $\Phi_{min} = -3606.6$, taken from Fig. 4.

## IV. CONCLUSIONS

1. It is found that with any self-consistent wave amplitude, trapped fast electrons quantitatively prevail over slow and fast passing particles. As the result with a strong nonlinearity, plasma consists of two antitropic beams: one of them consists of fast trapped electrons, and the other flow represents slow passing particles.

2. It is proved the discontinuity of distribution function when passing through the synchronism even in weak fields, as well as the vanishing of its derivative only at velocity $v = V$. These results conflict with the main conclusions of the quasi-linear theory (see, e.g., Ref. [4]). Namely, it also allows a distribution discontinuity, but only on the boundary surfaces $v = v_{1,2}(\psi)$, and the derivative $\partial f(v,\psi)/\partial v$ is considered to be zero not only on the plane of the synchronism, but also in its vicinity (plateau).

3. It is shown that waves with a nonzero average potential $\varphi_{av}$ can exist along with the Langmuir oscillations. Their phase velocity in contrast to the group velocity may be much greater than the values of $v_e$.

## ACKNOWLEDGMENT

We are grateful to Prof. Mikhail Bakunov for helpful comments and advice.

[1] I. B. Bernstein, J. M. Greene, and M. D. Kruskal, Phys. Rev. **108**, 546 (1957).
[2] P. S. Verma, S. Sengupta, and P. Kaw. Phys. Rev. E **86**, 016410 (2012).
[3] R. C. Davidson and Hong Qin, Phys. Rev. ST Accel. Beams **18**, 094201 (2015).
[4] A. I. Akhiezer, *Plasma Electrodynamics* (Pergamon Press, Oxford, 1975), Vol. **2**.
[5] L. A. Zelekson and N. S. Stepanov, Sov. Radiophys. **23**, 1046 (1980).
[6] V. G. Gavrilenko and L. A. Zelekson, Sov. J. Plasma Phys. **6**, 1227 (1980).

# Generation of antitropic electron beams by self-generated electric field. Kinetic description

Nickolay Stepanov and Lev Zelekson

This paper makes use of a one-dimensional kinetic model to investigate the nonlinear longitudinal dynamics of electron beams generated in the plasma under the influence of a self-generated electric field. It is expressed as where is a wave potential, , and charge particle distribution functions satisfy the Vlasov equation. It is proved that its correct solution is characterized by sudden change in the resonant part of the distribution function. Hence, in particular, the incorrectness of the established in the literature point of view follows that the fast, with velocities over V, and slow, with velocities under V, trapped particles are described by the same distribution function. Also for the first time it is shown that the self-generated strong electric field always produces antitropic electron beams with the velocities much larger than the value V, including the cold plasma limit. The possibility of implementing a new class of self-consistent wave structures with a nonzero average potential is shown. Maximum possible amplitude of a self-generated electric field in a stationary collisionless plasma is determined

**Categories**

Primary: Plasma Physics (physics.plasm-ph)

Cross lists: